# Author Once, Publish Everywhere: Portable Metadata Authoring with the CEDAR Embeddable Editor

Martin J. O'Connor, Marcos Martínez-Romero, Attila L. Egyedi, Mete U. Akdogan, Michael V. Dorf, and Mark A. Musen

Stanford Center for Biomedical Informatics Research
Stanford, CA 94304, U.S.A.
`martin.oconnor@stanford.edu`

## Abstract

High-quality, "rich" metadata are essential for making research data findable, interoperable, and reusable. The Center for Expanded Data Annotation and Retrieval (CEDAR) has long addressed this need by providing tools to design machine-actionable metadata templates that encode community standards in a computable form. To make these capabilities more accessible within real-world research workflows, we have developed the CEDAR Embeddable Editor (CEE)—a lightweight, interoperable Web Component that brings structured, standards-based metadata authoring directly into third-party platforms. The CEE dynamically renders metadata forms from machine-actionable templates and produces semantically rich metadata in JSON-LD format. It supports ontology-based value selection via the BioPortal ontology repository, and it includes external authority resolution for persistent identifiers such as ORCIDs for individuals and RORs for research organizations. Crucially, the CEE requires no custom user-interface development, allowing deployment across diverse platforms. The CEE has been successfully integrated into generalist scientific data repositories such as Dryad and the Open Science Framework, demonstrating its ability to support discipline-specific metadata creation. By supporting the embedding of metadata authoring within existing research environments, the CEE can facilitate the adoption of community standards and help improve metadata quality across scientific disciplines.

## Introduction

The FAIR Guiding Principles have become a cornerstone of modern data stewardship, emphasizing that data should be Findable, Accessible, Interoperable, and Reusable—not just for humans, but also for machines [1]. While these principles are broadly endorsed, operationalizing them remains a challenge, particularly for the metadata that must accompany scientific datasets to make the datasets FAIR in practice. The difficulty lies in the abstract nature of the FAIR principles themselves, which call for "rich" and

"community-relevant" metadata without specifying how those qualities should be formally represented or assessed.

The CEDAR Workbench [2] has addressed this challenge by offering a system for creating machine-actionable metadata templates that encode community standards in a structured and computable form. These templates define not only the required metadata fields but also the value types and controlled vocabularies appropriate for each field. Originally, users completed these templates using a centralized Web interface—the CEDAR Metadata Editor—hosted within the CEDAR platform. This environment ensured that metadata conformed to both the structural and semantic conventions of the relevant scientific community [3]. However, the centralized model required researchers to leave their native platforms and engage with an external tool, limiting its integration into routine workflows.

To address this limitation, we developed the CEDAR Embeddable Editor (CEE): a reusable, standards-compliant Web Component that brings CEDAR's metadata authoring capabilities directly into third-party applications. Even with a robust infrastructure for template design and management, metadata quality and standards compliance depend on ease of use and contextual integration. The CEE allows external systems to leverage CEDAR's structured templates without requiring users to leave their existing environments—embedding metadata creation within the flow of research and data submission.

Critically, CEE eliminates the need for platform developers to build or maintain custom user interfaces for metadata entry and display. Because the editor dynamically renders forms based on CEDAR templates and returns structured metadata in JSON-LD, the presentation and data-creation logic are entirely encapsulated within the component. This design ensures that any changes to metadata templates—whether structural updates, controlled vocabulary adjustments, or community-standard revisions—are automatically reflected in the rendered forms without further development effort. As a result, the CEE enables deployment of standards-compliant metadata interfaces across diverse platforms, while reducing the engineering overhead typically associated with metadata user interface development and maintenance.

This paper introduces the design and implementation of the CEDAR Embeddable Editor. It explains how the CEE fits into the broader CEDAR architecture, how it supports domain-specific metadata creation through template-driven rendering, and how it enables consistent, ontology-linked metadata output in JSON-LD format. In addition to ontology-based value selection, the CEE also supports external authority fields, allowing users to incorporate validated identifiers from registries such as ORCID [4], ROR [5], and the EPA

CompTox [6] system directly into metadata records. We present real-world deployments of the CEE in platforms such as the Open Science Framework [7] and Dryad [8], and show how embedding standards-aligned metadata tools directly into data publication workflows can promote wider adoption of community practices and improves metadata quality across disciplines.

## Related Work

A wide array of tools and frameworks have been developed to support the creation of metadata for research datasets, particularly in response to the increasing emphasis on the FAIR principles.

Several general-purpose tools, such as the Dublin Core Metadata Generator [9] and the DataCite Metadata Generator [10], offer lightweight interfaces for generating metadata compliant with high-level standards. While useful for basic dataset documentation, these tools are constrained to a single, highly abstract metadata model and lack support for the detailed, context-specific attributes required in many research domains. As a result, the metadata they produce are often too shallow and generic to support meaningful data discovery, interoperability, or machine-actionable reuse.

To address the limitations of generic tools, various domain-specific frameworks have emerged. Examples include ISA-Tab [11] for life sciences, and the Ecological Metadata Language (EML) [12] for environmental science. These systems embed richer semantics and structured fields aligned with disciplinary standards. However, they often presume substantial familiarity with community-specific ontologies and metadata practices. This learning curve can be a significant barrier, particularly for interdisciplinary users or researchers new to metadata curation. They also are tailored to their particular scientific disciplines and do not represent general solutions,

Many technically capable metadata systems are hampered by usability issues. Tools such as Metatool [13] and Morpho [14] illustrate this tension: While powerful, their complex interfaces and steep learning curves have limited broader adoption. Tools that do not prioritize intuitive interaction risk producing inconsistent or incomplete metadata, even when supported by sound metadata models.

Frameworks like FAIRsharing [15] and RO-Crate [16] have emerged to bridge the FAIR principles and implementation. FAIRsharing catalogs community standards and encourages reuse, while RO-Crate provides a packaging format for datasets and associated metadata. However, their adoption often requires substantial customization and technical overhead, limiting accessibility for many research groups. Moreover, while these efforts promote consistency, they do not directly address the problem of authoring rich metadata in real-world, embedded contexts.

Another critical limitation of current tools is their poor integration with external research platforms. Although systems such as OpenRefine [17] facilitate metadata enrichment and linking, they are often external to the platforms where data are managed and shared. Embedding them typically demands additional development effort and can introduce friction into data workflows. This integration gap limits the routine use of metadata tools at the point of data publication or sharing.

The CEDAR Workbench aims to address many of the longstanding challenges in metadata authoring by introducing machine-actionable templates that encode community standards and domain-specific requirements in a reusable, computable form. However, early CEDAR integrations with third-party systems required researchers to leave their native environments and use a centralized Web interface—the CEDAR Metadata Editor—for metadata entry. While this interface enabled form-based metadata authoring with ontology-based value selection and semantic validation, integration into external systems was cumbersome. Metadata needed to be authored on the CEDAR platform itself, and downstream submission workflows had to extract, transform, and relay data manually or via batch processes. For example, an early integration with the LINCS Consortium [18] demonstrated the feasibility of end-to-end CEDAR-facilitated metadata pipeline but also revealed the friction involved in connecting CEDAR-hosted metadata creation with external data repositories and submission systems.

The CEDAR Embeddable Editor (CEE) was developed to overcome these limitations by enabling CEDAR templates to be rendered directly within third-party applications. Delivered as a lightweight, standards-compliant Web Component, the CEE eliminates the need for external metadata portals or custom interface development. It dynamically adapts to template changes, supports ontology-based value recommendations, and outputs structured, metadata enriched with persistent identifiers from external authorities such as ORCID and ROR. Crucially, it requires no custom user interface development, allowing deployment across diverse platforms without the need for additional frontend engineering. This approach combines domain specificity, usability, and seamless integration—directly addressing the technical and adoption barriers encountered in earlier systems.

In summary, while existing metadata tools have advanced standardization and semantic expressivity, few offer the combination of rich domain modeling, intuitive design, and direct integration that the CEE enables.

# The CEDAR Embeddable Editor

To extend and improve the capabilities of the CEDAR platform, the team developed the CEDAR Embeddable Editor (CEE). Building on the template-driven infrastructure of the CEDAR Workbench, the CEE extends CEDAR's metadata creation capabilities into external platforms through a lightweight, self-contained Web Component. It enables structured, ontology-based metadata authoring directly within host applications, combining flexibility, semantic precision, and ease of deployment to support a wide range of scientific workflows.

## Background: Overview of the CEDAR Platform

The CEDAR (Center for Expanded Data Annotation and Retrieval) platform [2, 19] was created to address to a central challenge in scientific data stewardship: While the FAIR principles urge that metadata should be "rich" and compliant with community standards, few tools provide a practical means for researchers to achieve these goals [3]. The burden of interpreting abstract guidelines, selecting appropriate standards, and ensuring consistency across records typically falls on individual researchers or data curators—often without sufficient training, resources, or tooling. CEDAR was conceived to lower these barriers through automation, standardization, and intuitive user interfaces focused on community-driven design.

At the heart of the CEDAR platform is the concept of machine-actionable metadata templates [20]. These templates encode the structure, semantics, and possible expected values of metadata records using a formal, JSON Schema-based model. Unlike free-text reporting guidelines, which are often ambiguous and difficult to enforce, CEDAR templates make metadata specifications explicit and computable. Each template defines a collection of fields, their data types, required constraints, and associated controlled vocabularies drawn from community ontologies accessible through the BioPortal ontology repository [21]. CEDAR support a rich array of field types, including text, number, date/time, Boolean, checkboxes, image, and video fields. Fields can also be defined as single-valued or multi-valued. Templates may be assembled from reusable components—known as template elements—that encapsulate commonly used metadata patterns (e.g., specimen descriptors, instrumentation settings, study design attributes).

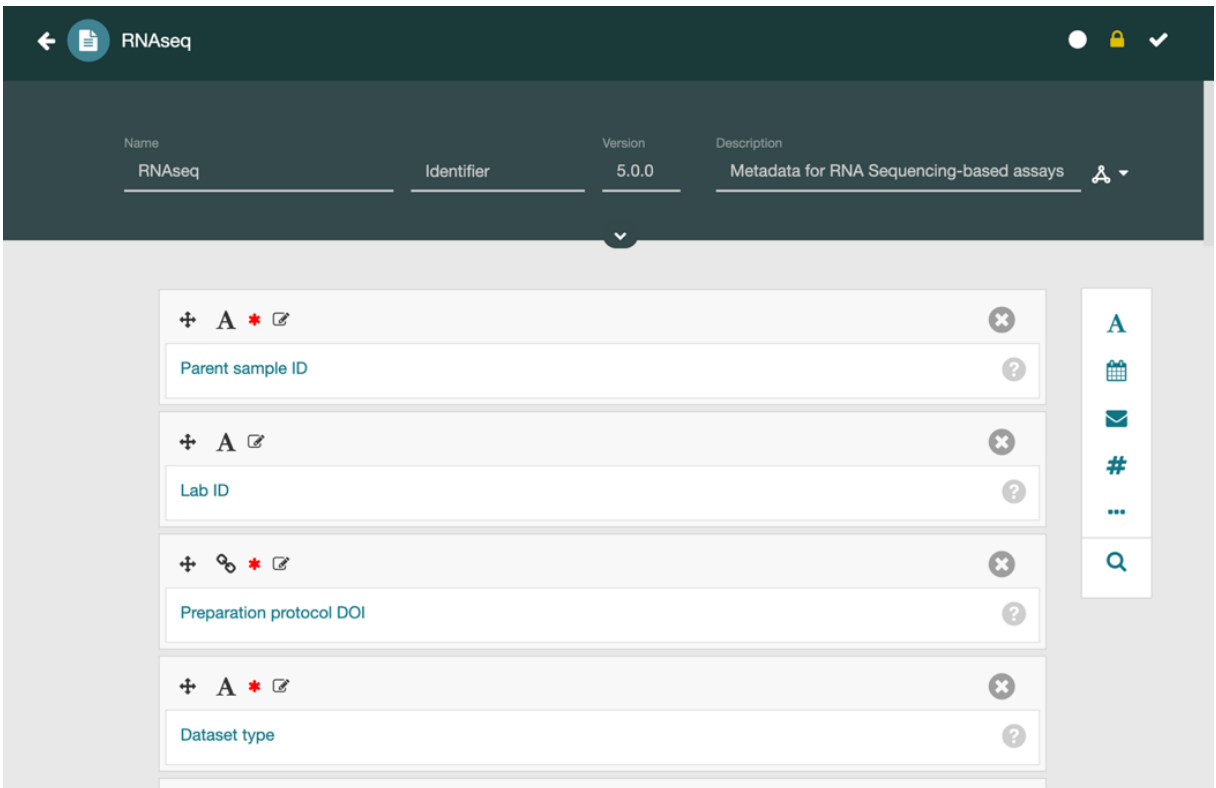

**Figure 1:** A CEDAR template in the Template Designer tool. Template authors can use the Template Designer to interactively build templates to describe metadata reporting guidelines. Here, the template describes metadata for an RNAseq-based assay.

The typical CEDAR metadata development workflow begins with template creation. Domain experts use CEDAR's Template Designer (Figure 1) to build templates that reflect the information needs of a specific scientific community, experimental method, or data-sharing requirement. Through a drag-and-drop interface, users define the structure of a metadata record without writing code.

Once a template is published, it becomes the blueprint for generating a user-facing metadata authoring interface. In the CEDAR Workbench, a tool called the Metadata Editor (Figure 2) can be used to automatically generate a Web-based form from a template's schema. Form fields are rendered according to their type (e.g., text, number, date, controlled term) and, if the fields are ontology-linked, users are guided to select appropriate ontology terms through drop-down autocompletion menus backed by BioPortal's ontology services. Constraints such as required fields, allowed values, or term hierarchies are automatically enforced by the editor, ensuring consistency and adherence to the underlying specification.

**Figure 2**. CEDAR-generated Web-based form illustrating how metadata can be acquired from users in the Metadata Editor tool. Here, the form is generated automatically from a template that specifies metadata for a RNAseq-based biomedical assay. In this case, the metadata author has completed four initial fields and is being presented with a set of choices for a field called Analyte class. Once completed, the user can save the completed form, whereupon it is stored in CEDAR.

As a user completes the form, CEDAR incrementally builds a structured metadata instance that conforms to the selected template. The resulting record is exported in JSON-LD [22], a linked data format that preserves the semantic structure of the metadata and allows for easy integration with downstream systems and repositories. This output format is key to ensuring the metadata's interoperability and machine-readability, as each data element is unambiguously linked to the vocabulary or ontology from which it was drawn.

Beyond template creation and instance authoring, the CEDAR Workbench also supports metadata sharing, collaboration, and validation. Templates and metadata instances can be versioned, published to community folders, or shared with specific collaborators. Additionally, CEDAR offers tools for metadata validation and data quality assessment, which help curators evaluate completeness and compliance before metadata are submitted to external repositories.

Major biomedical consortia, such as HuBMAP [23] and SenNet [24], are using CEDAR as the foundation of their entire metadata management workflow. Others, for example HEAL [25], have started integrating CEDAR as a central component of their metadata management processes. CEDAR is also incorporated in data portals, such as Health-RI, a national initiative in the Netherlands to improve the reuse of health data for policy, research, and innovation [26]. The RADx Data Hub [27], an initiative to create a national resource for managing COVID-19 data, uses technologies derived from the CEDAR platform to support its metadata management capabilities.

The CEDAR platform offers a comprehensive solution for defining, generating, and validating high-quality metadata. By transforming metadata standards into machine-actionable templates and automatically generating intuitive, standards-compliant interfaces, CEDAR empowers scientific communities to formalize and scale their metadata practices.

## The CEDAR Embeddable Editor

While the CEDAR Workbench provides a robust environment for designing and managing community-aligned metadata templates, its original deployment model required users to work within a dedicated platform. This separation often limited adoption, as researchers prefer to interact with metadata tools inside the systems that they already use to manage data. Platform developers, in turn, faced significant barriers in integrating external metadata tools, especially those requiring custom interface work or ongoing maintenance. The CEE addresses this gap by bringing CEDAR's full template-driven metadata capabilities directly into third-party environments—enabling structured, ontology-linked metadata to be authored and displayed within existing research platforms, without the need for bespoke user interface development.

### Component Architecture and Embedding

The CEE is implemented as a lightweight, standards-compliant Web Component designed for integration into a wide range of Web applications. It is distributed as a framework-agnostic package via `npm`, with no external dependencies. Once installed in an environment, the component can be invoked using the custom HTML tag `cedar-embeddable-editor`.

Because it conforms to the Web Components standard, the CEE works identically across modern frontend frameworks—including React, Angular, and Vue—as well as in static HTML contexts. Developers can instantiate it either directly in the DOM or within an HTML iframe, and supply the necessary inputs via custom parameters. Required inputs include a template (a CEDAR metadata template in JSON Schema format) and a configuration object, which controls language settings, display mode, and interface features. Optionally, an existing metadata instance may also be passed for editing or display.

This architecture allows the editor to be embedded in diverse platforms—from data repositories to mobile application interfaces—with minimal integration effort and without custom UI development.

### Template-Driven UI Generation

At the core of the CEE is its ability to interpret and render CEDAR templates. As mentioned, these templates may include complex nested elements and single- or multi-valued fields. Templates also accommodate display specifications such as field visibility conditions and language-specific labels via language maps. When a template is passed to the CEE, it is interpreted at runtime and dynamically rendered into a fully functional metadata form. Any changes to the template—such as updated ontology bindings or newly required fields—are immediately reflected in the rendered form, eliminating the need for redeployment or manual schema-to-UI synchronization.

### Metadata Acquisition and Output

The CEE guides users through metadata entry in real time, using template-driven validation to ensure conformity with structural and semantic requirements. Required fields are flagged, value types are enforced, and inconsistent or incomplete records are surfaced during authoring via contextual feedback. As users complete a form, the CEE incrementally constructs a metadata instance that conforms precisely to the supplied template.

As discussed, all output is serialized as JSON-LD [22], a linked data format that preserves the structure of the metadata and encodes controlled term values with their corresponding vocabulary term or ontology reference. Completed metadata instances can be retrieved using the component's API, stored locally, sent to a backend, submitted to the CEDAR repository, or exported for downstream use. The use of JSON-LD ensures that resulting metadata adhere to a standard representation, are machine-readable, and can be easily processed by downstream systems, including analytic pipelines and data catalogs (Figure 3).

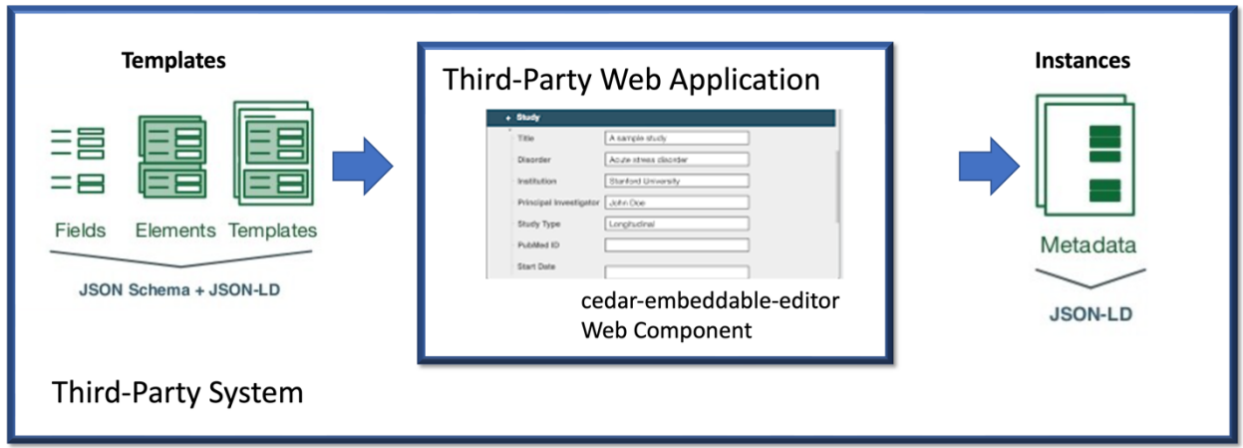

**Figure 3.** This diagram illustrates the integration workflow of the CEDAR Embeddable Editor (CEE) within a third-party Web application. On the left, domain-specific metadata templates—composed of fields, elements, and templates defined using JSON Schema and JSON-LD—are created in a third-party system. These templates are rendered in the central component using the `cedar-embeddable-editor` Web Component, which allows users to input structured metadata directly within the host application. On the right, completed metadata instances are generated in JSON-LD format, ready for storage, reuse, or publication.

## Semantic Integration

A key feature of the CEE is its built-in support for ontology-based value selection. Fields can be linked to ontologies, ontology branches, and value sets hosted on the BioPortal ontology repository [21]. During metadata entry, the editor suggests relevant terms—displayed with labels and synonyms—sourced from the configured ontologies and value sets. When a user selects a term, the editor records both its human-readable label and its IRI, ensuring semantic precision and enabling downstream reuse.

The CEE also supports integration with external authority systems for persistent identifier fields (Figure 4). Currently supported authorities include ORCID, for identifying individual researchers, and the Research Organization Registry (ROR), for identifying research organizations. We have recently added the U.S. Environmental Protection Agency CompTox API as an authority for selecting per- and polyfluoroalkyl substances (PFAS)—a class of synthetic chemicals widely studied in environmental science and toxicology. This integration enables researchers to link PFAS-related metadata fields to precise chemical identifiers maintained by the EPA, supporting more accurate reporting and downstream reuse in environmental datasets.

For these fields, the CEE performs REST-based searches against external authority services to autocomplete user input and to validate selections. These requests are routed through an intermediating microservice provided by the CEDAR system, which facilitates communication with multiple external APIs by handling authentication, standardizing request formats, and normalizing responses. When users select from the returned options, the CEE records globally resolvable, authoritative identifiers. Additional external authority fields are planned for future releases of the CEE.

## Interaction Modes and Accessibility

The editor supports multiple operational modes to accommodate different usage scenarios:

1. Form Entry Mode: The default mode, used for creating new metadata instances.
2. Edit Mode: Allows modification of a previously authored metadata instance.
3. View-Only Mode: Displays a template or instance without permitting edits—useful for audit trails, data reviews, and read-only contexts such as metadata visualization platforms.

The editor integrates visually with its host environment by inheriting local styles, maintaining a consistent user interface. Accessibility is a core feature: the CEE supports complete keyboard operability and screen reader compatibility via standardized WAI-ARIA role annotations [29].

The CEE ships with built-in support for multiple languages. Language selection is configured via the initialization object passed to the editor. Developers can add new translations by mirroring the structure of the provided English language map and updating the configuration. The editor gracefully falls back through available language maps when keys are missing, with diagnostics surfaced through a debug trace.

**Figure 4.** This image shows a metadata entry form rendered by the CEDAR Embeddable Editor (CEE), with the Principal Investigator field completed using a valid ORCID identifier. Below, the Institution Research Organization Registry (ROR) field shows the CEE's autocomplete capability, allowing users to search and select standardized institutional identifiers. This external authority integration ensures precise, interoperable metadata by linking to authoritative external registries.

To support accurate metadata entry, the editor provides real-time validation. Color-coded prompts highlight missing required fields, datatype mismatches, and ontology term issues as the user interacts with the form. A collapsible validation ribbon summarizes all outstanding problems in a centralized view, enabling users to resolve issues before submission. For more comprehensive assessments, the CEE can generate structured data quality reports based on the template specification—flagging missing values, unselected controlled terms, or inconsistencies across the metadata record.

## Deployment and Configuration

The CEE is designed for low-friction deployment. Because the UI is dynamically rendered based on centrally managed templates, updates to metadata standards—such as revised field definitions or updated value sets—are automatically reflected in the editor without any code changes to the host application.

The editor's configuration object—passed as a JavaScript property or bound to the HTML tag—exposes fine-grained control over appearance and behavior, including:

1. Template and instance injection
2. Form mode (edit/view-only)
3. Language and localization paths
4. Display of headers, footers, and validation ribbons
5. Styling integration via scoped CSS classes and slot-based layout hooks

Templates used by the CEE are initially authored within the CEDAR Workbench by the third-party deployer—typically a developer, data manager, or community standards lead responsible for metadata integration within a given platform. The deployer uses the Workbench's visual editor to construct machine-actionable metadata templates, specifying field structure, cardinality, validation constraints, controlled vocabulary bindings, and language-localized labels. Once finalized, each template can be exported from CEDAR as a standalone JSON Schema document. Rather than relying on CEDAR to serve these templates at runtime, most third-party systems store and manage them locally—either embedding the JSON directly in their application logic or maintaining a versioned template registry within their infrastructure.

At runtime, the CEE receives the JSON Schema template as a property or API-delivered object and renders the corresponding metadata form on the client side. This design decouples form generation from CEDAR's live infrastructure, allowing the third-party platform to retain full control over which template versions are in use and when updates are deployed. The deployer is thus responsible not only for designing the template but also for maintaining its lifecycle—tracking changes, managing backward compatibility, and aligning with evolving community standards or platform requirements.

Similarly, all metadata instances authored using the CEE are retained and managed entirely by the host platform. As users complete and submit forms, the resulting metadata—structured as JSON-LD and semantically enriched with persistent identifiers and ontology-linked terms—are returned to the host application via the CEE's API. From there, the third-party system is responsible for storing, indexing, validating, and linking these metadata records with associated digital objects (e.g., datasets, software packages, protocols). CEDAR does not store or archive metadata created through the CEE unless explicitly configured to do so. This model helps ensure that integrations are lightweight and that metadata ownership, persistence, and discoverability remain under the control of the embedding platform.

The CEE supports styling integration through a set of scoped class names applied to UI elements, allowing developers to override fonts, spacing, or color schemes using host-defined CSS. Additionally, the component exposes elements that enable custom theming or layout control within the host application's design system. These features ensure that the editor blends visually with surrounding components and remains consistent across embedding environments.

The component has been deployed in environments built using React, Django, and custom JavaScript frameworks, and tested across all major browsers and a selection of mobile devices.

## Results

The CEE's lightweight deployment model and framework-agnostic architecture have made it suitable for integration across a variety of research platforms. Its ability to render rich, standards-based metadata forms without requiring platform-specific interface development has enabled adoption in diverse environments. These deployments demonstrate how the editor supports community-specific metadata practices and enables semantic interoperability without disrupting existing workflows.

**Figure 5.** This OSF interface presents users with a selection of metadata templates defined using CEDAR. Each template—such as those for Psych-DS, human cognitive neuroscience data, or generic datasets—encodes a community- or domain-specific metadata standard. Upon selecting a template, users are presented with a structured metadata entry form rendered by the CEDAR Embeddable Editor (CEE), enabling guided, standards-compliant metadata creation directly within the OSF platform.

Within the Open Science Framework (OSF) [30], the CEE is embedded into the metadata authoring workflow during project registration and dataset submission [7]. As part of an initiative by the Center for Open Science to improve metadata quality and field-specific coverage, OSF first prompts researchers to complete general metadata describing the overall study—such as title, contributors, funding, and study purpose—before selecting a discipline-specific metadata template (Figure 5). Upon this selection, the CEE is invoked to render the corresponding structured form within the OSF interface. These templates, designed in collaboration with domain experts and curated in the CEDAR Workbench, include discipline-specific schemas for cognitive neuroscience, clinical trials, ecology, and other areas of active research. Each template encodes not only the fields required for rich documentation of study context and methods but also includes bindings to relevant ontologies through BioPortal—for example, recommending terms from the Cognitive Atlas or MeSH depending on the selected domain.

**Figure 6.** This screenshot shows the Psych-DS template rendered within the Open Science Framework (OSF) using the CEDAR Embeddable Editor (CEE). The form provides a structured interface for entering metadata fields defined by the Psych-DS metadata standard. By embedding the CEE, OSF enables users to create semantically rich, standards-compliant metadata records directly within their research workflows.

By embedding these templates directly through the CEE, OSF provides users with structured guidance that encourages the entry of consistent, semantically valid metadata. Users do not need to leave the OSF environment or install additional tools. The editor enforces constraints and offers intelligent autocompletion for ontology-linked fields in real time. The resulting JSON-LD output ensures that the metadata can be programmatically analyzed, reused by other platforms, or federated with disciplinary metadata repositories. This integration also lays the foundation for more granular, scalable metadata capture across OSF-hosted projects. By decoupling template logic from the platform's user interface code, OSF can support ongoing expansion of template coverage across fields without requiring frontend redevelopment—enabling a long-term strategy for increasing metadata richness and FAIR compliance across diverse scientific disciplines.

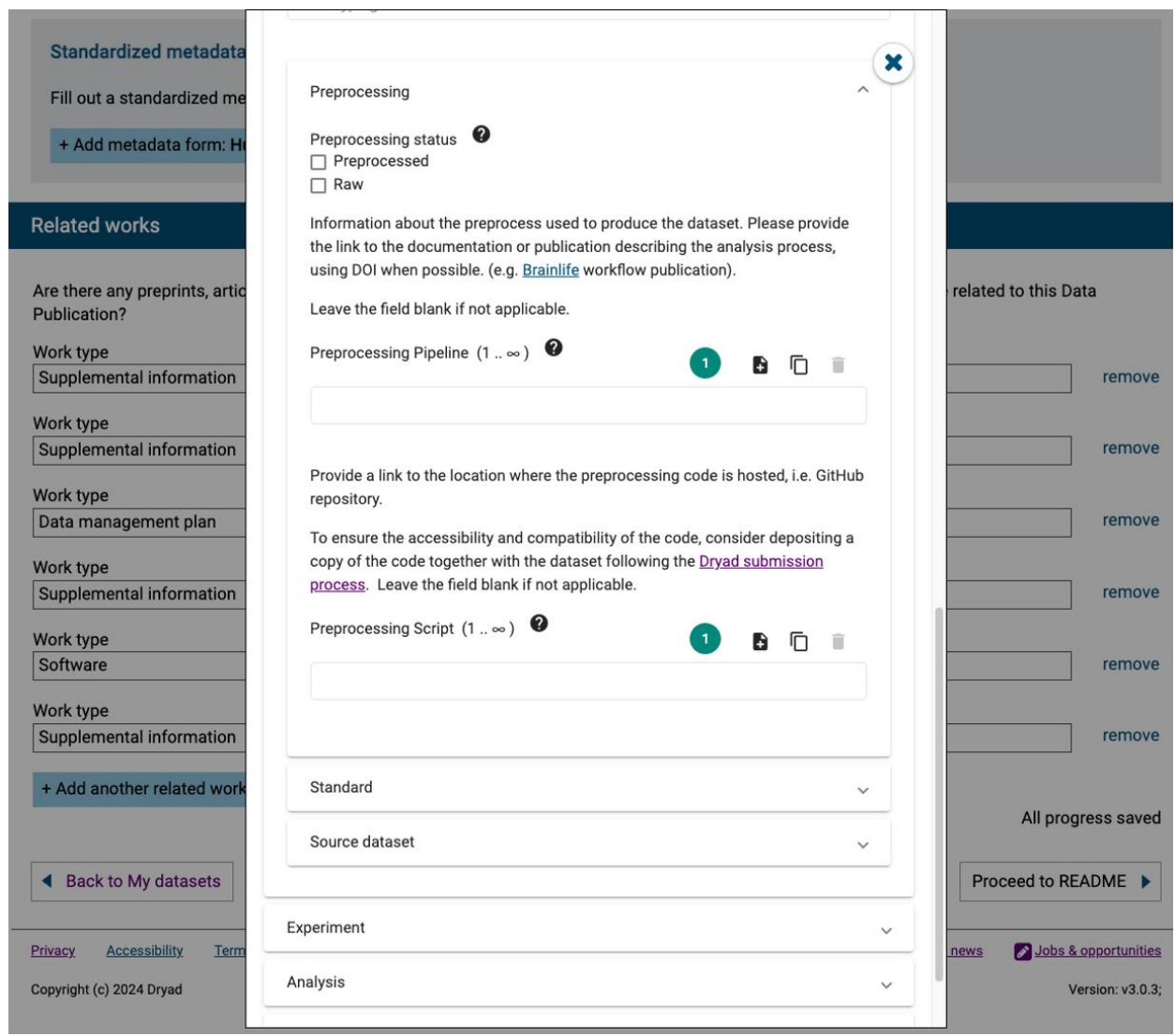

**Figure 7.** This screenshot shows a neuroscience metadata template rendered within the Dryad data repository using the CEDAR Embeddable Editor (CEE). The displayed section captures structured metadata related to data preprocessing, including status, pipeline references, and script locations. By embedding CEE, Dryad enables researchers to input detailed, standards-based metadata directly within the dataset submission workflow, supporting reproducibility and alignment with open data best practices.

Dryad's [31] deployment of the CEE focuses on improving metadata quality across multiple domains, including neuroscience and high-throughput biology. In partnership with domain experts and CEDAR developers, Dryad first introduced a neuroscience-specific metadata template to capture detailed contextual information essential for reuse (Figure 7). This template includes fields for experimental paradigm, participant demographics, imaging modality, cognitive domain, and data acquisition protocols, many of which are linked to controlled vocabularies such as the Cognitive Atlas and OpenNeuro Vocabulary via BioPortal. More recently, Dryad also integrated a set of metadata templates developed by the HuBMAP Consortium [32, 33], which describe assay workflows, sample characteristics, and tissue metadata for cellular-resolution datasets. These templates allow Dryad to support metadata capture for a broader range of biological submissions. During dataset submission, users are prompted with CEDAR-rendered forms that reflect the selected template. The editor enforces constraints, provides ontology-based term recommendations in real time, and exports completed records as JSON-LD. The resulting metadata are stored by Dryad alongside the associated dataset, making them available for indexing, discovery, and downstream reuse. Dryad staff noted that these integrations lowered the barrier to submitting specialized domain-specific data, helping to align metadata practices with the FAIR principles and community standards [8].

The RADx and HuBMAP programs use the CEE in a read-only mode to display templates and metadata authored externally but rendered for user-friendly inspection. In these contexts, the CEE provides a consistent interface for reviewing metadata associated with complex biological datasets. The RADx Data Hub [27], for instance, uses the CEE to present metadata records for COVID-19 diagnostic projects. The visual rendering of nested, structured information improves readability. In HuBMAP, the CEE serves as the authoritative public interface for viewing all metadata templates used across the consortium [32, 33]. Each template is rendered in a readable, structured format directly from its machine-actionable specification, enabling researchers, tool developers, and reviewers to inspect the exact metadata requirements in a standardized, accessible form.

A unifying feature across all CEE deployments is the use of JSON-LD as the output format for metadata instances. Regardless of the domain or platform, metadata authored through the CEE are encoded in a semantically structured, machine-readable format that aligns with linked data principles. This common representation enables metadata to be validated, indexed, and transformed using standard Web-based tools, while preserving field-level semantics and ontology references. Platform developers across OSF, Dryad, and HuBMAP emphasized the value of this consistency—not only for immediate use in submission pipelines and UI display, but also for long-term goals such as metadata federation, reuse across repositories, and automated schema evolution. JSON-LD also facilitates integration with persistent identifier systems (e.g., ORCID, ROR) and supports downstream applications including DOI registration, search indexing, and metadata enrichment via external services.

Together, these deployment experiences demonstrate that the CEE can lower the barrier to structured metadata authoring and provides a practical path for extending FAIR practices across a wide range of research infrastructures.

# Discussion

The CEDAR Embeddable Editor brings structured, standards-based metadata authoring directly into third-party platforms through a lightweight, framework-agnostic Web Component. It dynamically renders forms from machine-actionable templates, enforces community-defined validation rules, supports ontology-based value selection and external identifier resolution, and outputs metadata in semantically rich JSON-LD format. By encapsulating all metadata logic within the component itself, the CEE eliminates the need for custom interface development while ensuring consistency with evolving metadata standards.

The adoption and integration of the CEE across multiple platforms underscores the growing value placed on metadata solutions that combine domain-specific adherence to standards with flexible deployment options. The CEE demonstrates that it is possible to deliver structured, standards-aligned interfaces for authoring metadata that are both technically robust and easy to embed within existing research workflows. By leveraging CEDAR's machine-actionable templates and its ontology infrastructure, the editor makes high-quality metadata entry accessible to researchers without requiring them to leave the environments in which they already manage and publish their work.

Compared to traditional metadata tools—whether general-purpose form builders or discipline-specific systems such as ISA-Tab [11] or EML [12] —the CEE strikes a unique balance between expressiveness, usability, and integration. Generic tools often lack the schema flexibility or ontology linkage required to support rich metadata with appropriate persistent identifiers, whereas more powerful domain-specific tools tend to be standalone applications with steep learning curves and limited adaptability. The CEE's lightweight, component-based architecture offers a middle path, enabling host platforms to support multiple community-defined schemas with minimal development burden. Its "no UI coding" paradigm lowers the barrier for third-party integration, allowing developers to rely on centrally managed templates that evolve independently of platform-specific user interfaces.

The architecture of the CEE highlights a broader design principle applicable well beyond metadata: decoupling form logic from platform code enables dynamic, standards-driven interfaces that can evolve independently of the systems embedding them. In any domain where users complete structured forms—whether for scientific metadata, grant applications, clinical records, or regulatory submissions—this separation allows user interfaces to reflect updated specifications, validation rules, and vocabularies without requiring software redeployment. The CEE's use of centrally defined, machine-actionable templates and its runtime rendering model illustrate how form-filling systems can become both more maintainable and more responsive to changing community needs. By outputting structured, semantically rich data—in this case, JSON-LD—the editor ensures compatibility with downstream processing, indexing, and integration tasks. This approach points to a more generalizable model for building intelligent, standards-aware form interfaces across a wide range of applications.

The deployment of CEDAR templates developed by the HuBMAP Consortium across both Dryad and the Open Science Framework (OSF) further illustrates the portability and ecosystem value of the CEE model. In collaboration with the HuBMAP metadata team, a set of ten templates—primarily covering assay descriptions, sample metadata, and tissue characterization workflows—were authored in the CEDAR Workbench using domain-specific schema elements and controlled vocabulary bindings [32]. These templates, originally designed to capture rich metadata for high-throughput biology experiments, have being integrated into both Dryad and OSF, where they are used to guide metadata entry during dataset submission and project registration. Notably, the same set of exported JSON Schema templates is being reused in both platforms without modification, with each system embedding the CEE and applying its own local configuration and styling.

These deployments demonstrate a core strength of the CEE architecture: Templates are fully portable and decoupled from any specific implementation or backend. Because the editor renders forms dynamically from JSON Schema templates at runtime and does not rely on server-side logic or templating engines, a single template can be reused across multiple platforms with no duplication of development effort. For research communities, this means that a well-defined metadata standard can be expressed once—using CEDAR's authoring tools—and then embedded consistently across the tools their researchers already use [3]. For platform developers, it enables schema alignment across systems while preserving autonomy over data flow, storage, and integration. As the HuBMAP deployment shows, shared metadata standards can now be operationalized across generalist and domain-specific infrastructure alike, promoting both metadata harmonization and localized control. This portability also creates new opportunities for maintaining authoritative, version-controlled template libraries that span multiple services and disciplines.

The results across deployments highlight that embedding structured metadata authoring into research workflows leads to more complete, consistent, and semantically valid metadata. The OSF and Dryad integrations show how the CEE can adapt to domain-specific needs while maintaining a unified implementation model. Meanwhile, the use of the CEE in read-only mode by RADx and HuBMAP illustrates the value of a consistent rendering layer for metadata review and quality control. Across these deployments, the feedback from researchers and platform administrators consistently emphasizes improved metadata quality, enhanced user experience, and reduced maintenance costs.

The CEE architecture enables a clean separation between metadata governance and platform implementation. Templates are defined and versioned within the CEDAR Workbench by community curators, standards developers, or platform-specific leads, while the embedding platforms remain responsible for how and when those templates are integrated into production workflows. This separation allows domain experts to evolve metadata standards over time—adding fields, adjusting validation logic, or updating ontology bindings—without requiring downstream systems to redevelop their user interfaces. In effect, the CEE makes it possible to delegate stewardship of semantic and structural metadata rules to the appropriate knowledge holders, while preserving implementation independence and stability for platform developers. This distributed governance model aligns with real-world research infrastructure, where metadata policies are negotiated across institutions, consortia, and sponsors, but must be supported by a diverse array of technical systems.

While the FAIR principles have become a cornerstone of research data policy, their implementation often remains aspirational, especially in domains lacking dedicated tooling. The CEE offers a concrete mechanism for operationalizing core aspects of FAIR by embedding machine-actionable metadata logic directly into submission workflows. Through real-time validation, persistent identifier resolution (e.g., ORCID, ROR), and ontology-based field constraints, the CEE can help ensure that the metadata researchers submit are not only structured and complete, but semantically meaningful and machine-interpretable. Because these features are encoded in reusable templates and automatically enforced by the editor, platforms can raise the metadata quality floor without burdening researchers with additional training or manual standards compliance. In this sense, the CEE shifts FAIR from a set of external recommendations into an embedded capability—making adherence to best practices a natural outcome of the submission process.

The CEE also serves as an effective bridge between general-purpose data platforms and the more specialized needs of disciplinary repositories. Many open data environments face a tension between broad accessibility and domain specificity: generalist platforms must support heterogeneous users and content types, while domain repositories demand high-quality, semantically rich metadata. By embedding the CEE and adopting discipline-specific templates, generalist platforms such as Dryad and OSF can support structured metadata capture without fragmenting their infrastructure or duplicating interface logic. This capability allows researchers to submit community-aligned metadata within familiar environments, while ensuring that those records remain compatible with downstream curation, indexing, or federation workflows in specialized ecosystems. As more research platforms converge on shared template libraries, the CEE provides a scalable mechanism for harmonizing metadata practices across the broader research data landscape.

Nevertheless, there are limitations. While the editor supports complex schemas, some scientific domains may still require capabilities not yet expressible in CEDAR's current metadata model. For example, CEE does not currently support advanced conditional logic such as dynamic field branching or question-skipping behavior based on prior responses—functionality that is often critical in clinical studies, environmental assessments, and longitudinal data collection instruments. Similarly, cross-field validation rules that involve dependencies across multiple metadata fields are not yet supported. These features are planned for future development, along with more expressive constraint languages and enhanced form logic.

Additionally, while the editor provides robust form-based entry, some users prefer working in tabular environments—particularly for bulk metadata entry or for records that naturally align with spreadsheet structures. Integration with spreadsheet-based metadata-entry tools, along with support for round-tripping between structured templates and spreadsheet formats, is an area of ongoing work [28]. Another promising direction is the automated extraction of metadata from primary data files (e.g., imaging headers, instrument logs), which could prepopulate template fields and reduce manual burden while preserving structured output. Addressing these limitations will help extend the applicability of the CEE to an even broader set of metadata use cases and scientific workflows.

By aligning a flexible technical architecture with a governance model for community-developed templates, the CEE offers a practical approach for how structured, standards-based metadata creation can be embedded across the research ecosystem. As adoption grows, we anticipate that the CEE will contribute not only to the quality and consistency of metadata, but also to the normalization of FAIR practices in everyday research workflows.

# Conclusions

The CEDAR Embeddable Editor is a lightweight, interoperable Web Component that brings standards-based, ontology-linked metadata authoring and visualization directly into third-party platforms. Developed within the CEDAR Workbench ecosystem, the CEE leverages machine-actionable metadata templates to render structured, semantically rich metadata forms without requiring custom user interface code. The editor has been integrated into a variety of platforms such as Dryad and the Open Science Framework, supporting both Web and mobile environments. By adhering to the same JSON-LD-based metadata model as CEDAR, the CEE ensures consistency across systems and automatically reflects updates to metadata specifications. Underpinned by the broader CEDAR architecture—which includes tools for template design, metadata validation, ontology integration via BioPortal, and spreadsheet interoperability—the CEE enables scientific communities to encode and enforce metadata reporting guidelines in a reusable format. It enables these communities to deliver high-quality metadata tools with reduced engineering effort—supporting the broader goal of making FAIR metadata the default rather than the exception.

# Data Accessibility Statement

The CEDAR Workbench and the CEDAR Embeddable Editor (CEE) source code is publicly available under the BSD 2-Clause License and maintained by the Stanford Center for Biomedical Informatics Research. The CEE software is available for execution at [https://github.com/metadatacenter/cedar-embeddable-editor](https://github.com/metadatacenter/cedar-embeddable-editor). All CEDAR Workbench source code is available from the Metadata Center landing page at [https://github.com/metadatacenter](https://github.com/metadatacenter). Metadata templates referenced in the Dryad and OSF deployments are available via the respective platform repositories or by request. All deployments use publicly available ontologies and value sets accessible through the BioPortal ontology repository at [https://bioportal.bioontology.org](https://bioportal.bioontology.org).

# Declarations

## Competing Interests

The authors declare no competing interests.

## Funding

This work was supported in part by Grant 2134956 from the National Science Foundation, by a grant from the Templeton World Charitable Foundation, and by Award OT2 DB000009 and Grant U24 GM143402 from the National Institutes of Health.

## Ethics Statement

This research did not involve human participants, personal data, or animal subjects and therefore did not require institutional ethical review.

## Author Contributions

MJO'C and MMR led the conceptualization and development of the CEE. ALE, MUA, and MVD contributed to deployment, implementation, and architecture. MAM supervised the project. All authors reviewed and approved the final manuscript.

## AI Tool Use Disclosure

Generative AI tools were used to assist with grammar refinement and to provide editorial feedback on structure and clarity. No content was generated autonomously by AI tools; all substantive contributions were made by the authors.